# Pressure-induced s,p-d electron redistribution accompanies structural transformation in amorphous Zr-Cu alloy under compression

Przemyslaw Dziegielewski*
0000-0002-7544-3960
Warsaw University of Technology Faculty of Physics Warsaw PL
przemyslaw.dziegielewski@pw.edu.pl
Jerzy Antonowicz
0000-0002-7781-7540
Warsaw University of Technology Faculty of Physics Warsaw PL
Konstantinos Georgarakis
0000-0003-0918-7310
Cranfield University Faculty of Engineering and Applied Science Cranfield GB
Oliver Lord
0000-0003-0563-1293
The University of Bristol School of Earth Sciences Bristol GB
Monica Amboage
0000-0002-6058-8446
Diamond Light Source Didcot GB
Dominik Daisenberger
0000-0001-5311-4747
Diamond Light Source Didcot GB
Lewis Clough
0000-0001-7296-1394
University of Edinburgh Edinburgh GB
Tetsuo Irifune
0000-0001-9194-6341
Ehime University: Matsuyama , Ehime, JP
Toru Shinmei
0000-0003-4980-060X
Ehime University: Matsuyama, Ehime, JP

**ABSTRACT** Extreme compression can reorganize atomic valence states, in some regimes driving charge from spatially extended orbitals into more compact ones. We investigate this effect in amorphous $Zr_{67}Cu_{33}$ metallic glass by combining high-pressure X-ray absorption fine structure measurements up to 69 GPa with molecular dynamics simulations and density functional theory calculations. The Zr K-edge XANES spectra show a pronounced increase in the pre-edge intensity and a change in its pressure dependence, whereas the Cu response is markedly weaker. Molecular-dynamics simulations reproduce the measured EXAFS spectra and provide reliable input for the electronic-structure calculations. The calculated Mulliken populations reveal pressure-induced depletion of Zr s and p-states, accompanied by an increase of the d-state population, while the density of states indicates enhanced p-d hybridization under compression. These results provide experimental and computational evidence for pressure-induced s,p→d electronic redistribution in zirconium and link this effect to the previously reported anomalous structural evolution of the compressed amorphous alloy.

## I. INTRODUCTION

Extreme pressures, exceeding tens of gigapascals, establish conditions under which the fundamental principles of chemistry and solid-state physics are substantially modified [1]. One theoretical framework describing compression-induced changes in the electronic structure of atoms is the concept proposed by Martin Rahm and coworkers [2,3]. They postulate that the isotropic compression of an atom leads to a fundamental violation of the Madelung rule in favor of a "hydrogenic" ordering, where the energy is primarily governed by the principal quantum number.

The underlying physical mechanism is the differential response of orbitals with varying azimuthal quantum numbers to spatial confinement. The s-type orbitals, lacking orbital angular momentum, exhibit the greatest radial extent; consequently, as pressure increases and atomic volume reduces, the energetic cost of confinement rises rapidly, particularly for s-electrons in metals. This follows directly from the Heisenberg uncertainty principle, as spatial confinement of the electron necessitates an increase in momentum uncertainty, translating into higher kinetic energy. In contrast to s-orbitals, d and f orbitals are inherently more compact and effectively shielded by inner electrons, causing their energies to increase significantly more slowly during compression. The p orbitals, with intermediate radial extent, are also destabilised but to a lesser degree than s-states.

*Contact author: przemyslaw.dziegielewski@pw.edu.pl

This energetic disparity leads to a reconfiguration of energy levels: orbitals with higher angular momentum (p, d, f) become more energetically favorable than s-orbitals. This results in a forced transfer of electrons from spatially extended shells to more compact ones (e.g., s→p, s→d, or p→d transitions). For instance, under pressure, lithium undergoes an s→p transition, effectively shifting its chemical character from that of an alkali metal to a p-block element. According to Rahm's model, these changes are not gradual but manifest as discrete electronic reconfigurations accompanied by drastic shifts in atomic parameters: a sharp decrease in the van der Waals radius and step-wise changes in electronegativity (χ). For zirconium, an element with a $[Kr]5s^2 4d^2$ configuration, the model predicts two reconfiguration thresholds: the first at approximately 2 GPa ($5s^2 4d^2 \rightarrow 5s^1 4d^3$) and a second, critical threshold at about 54 GPa, where the 5s orbital is completely depopulated ($5s^0 4d^4$). This transition is associated with an electronegativity jump of $\Delta\chi \approx 3.5$ eV/e, which substantially modifies the electronic character of Zr.

Although these predictions are based on modeling an isolated atom in an inert, neon-like environment [4], they can be extended to complex multicomponent systems. To date, the application of this framework to the solid state has been limited to a few crystalline compounds, in which it has successfully elucidated specific pressure-induced phenomena [5–7]. Metallic glasses (MG), however, constitute particularly suitable systems for assessment of the theory. The absence of long-range periodicity allows for the observation of polyamorphism - glass-to-glass phase transitions induced by electronic changes - free from the constraints of crystalline symmetry [8–10]. Numerous molecular dynamics (MD) simulations have reported an anomalous splitting of the first Zr-Zr coordination shell at pressures exceeding 50 GPa [11–14]. Atomic pairs of Zr exhibit distances that are nearly 25% shorter than the average, imparting a unique bimodal topology to the glass. In crystalline Zr, a series of polymorphic transformations has also been experimentally observed, with a significant transition occurring at approximately 50 GPa - strikingly close to the predicted pressure for the electronic configuration change in Zr [15–17], although a subsequent study has attributed these findings to experimental artifacts [18].

In this paper, we investigate the origin of high-pressure electronic reconfiguration in $Zr_{67}Cu_{33}$ MG by combining X-ray absorption fine structure (XAFS) spectroscopy with molecular dynamics (MD) simulations and calculations based on density functional theory (DFT). Through a systematic analysis of experimental and computational data, we identify pressure-induced redistribution of Zr valence charge. These phenomena are interpreted within the framework of Rahm's predictions, providing a test of that model on amorphous metallic alloys under extreme compression.

## II. METHODS

### A. XAFS experiment

XAFS measurements were conducted at beamline I18 [19], Diamond Light Source, on a $Zr_{67}Cu_{33}$ metallic glass foil loaded by a Ne pressure-transmitting medium in a diamond anvil cell (DAC) equipped with nanopolycrystalline anvils [20]. Pressure was increased up to 69 GPa and determined from the volume of a Au calibrant measured by XRD and ruby fluorescence spectroscopy, with an uncertainty of 1–3 GPa. Both Cu (8979 eV) and Zr (17998 eV) K-edge spectra were measured at each P point. Detailed experimental procedures and pressure calibration methods are provided in the Supplemental Material (SM).

### B. Calculations

#### *1. Molecular dynamics*

MD simulations were performed using LAMMPS [21] for a $Zr_{67}Cu_{33}$ system containing 43904 atoms placed in a cubic simulation box with periodic boundary conditions. Interatomic interactions were calculated using the Sheng EAM-type potential [22,23], parameterized against a first-principles potential-energy surface that includes equations of state, lattice dynamics, adiabatic elastic constants (including a pressure correction), and liquid-state properties. It is worth noting that the electronic properties in the EAM potential derive from the DFT calculations used in its development, while electrons are not explicitly treated in MD simulations. The simulations were carried out in the NPT ensemble. The initial crystalline configuration was pre-relaxed, heated, and annealed at 2000 K, then quenched to 300 K at a cooling rate of $10^{12}$ K/s. The resulting glass was subsequently compressed up to 100 GPa at a rate of 0.1 GPa/ps. Using the same protocol, a classical MD simulation was also performed on a 128-atom system,

*Contact author: przemyslaw.dziegielewski@pw.edu.pl

which served as the configuration for the DFT calculations (see SM for details).

### 2. *DFT calculations*

DFT calculations were performed using FHI-aims [24] with an atom-centered basis of numerical atomic orbitals. Representative atomic snapshots for DFT analysis were extracted from MD trajectories at 5 GPa intervals from 0 to 100 GPa. Electronic-structure quantities (DOS, Mulliken charges) were computed using a 2×2×2 k-point grid using the PBEsol exchange-correlation functional. Given the significance of inner-shell electrons, an "intermediate" basis set was used (see SM).

### 3. *Breakpoint analysis*

To identify pressure-induced changes in slope for a given observable, we fitted a continuous piecewise-linear model with two segments meeting at a breakpoint, following standard practice for detecting changes in slope in noisy experimental or simulated data. Breakpoint locations and 95% confidence intervals were obtained via residual bootstrap (2000 resamples), following established practice for quantifying breakpoint location uncertainty in piecewise-linear regression [25,26]. Methodological details are given in the SM.

## III. RESULTS

### A. XANES analysis

The primary experimental evidence for electronic structure modifications in the $Zr_{67}Cu_{33}$ alloy under extreme compression is provided by changes in the X-ray absorption near-edge structure (XANES) spectra. These spectra directly probe the density of unoccupied states above the Fermi level ($E_F$), and their pressure evolution reflects modifications in both the atomic potential and orbital hybridization [27–31].

Across the examined pressure range (0-69 GPa), we observe a systematic shift of the absorption edge to higher energies. For Cu, this shift reaches 1.1 eV (FIG. 1(a) and (c)), whereas for Zr it is higher, reaching 1.6 eV (FIG. 1(b) and (c)). This pronounced blue shift is fundamentally attributed to the increase in the kinetic energy of electrons confined within the decreasing atomic volume, as expected from consideration of the Heisenberg uncertainty principle, which raises the energy of excited states. However, it is the evolution of the spectral shape, rather than the edge shift itself, that provides critical insight into the electronic transformation.

In amorphous Zr-Cu alloys, the XANES spectra of Zr and Cu typically exhibit a shoulder or kink on the main rising absorption edge, often referred to in the literature as a "pre-edge" feature [29,32,33]. For Zr, this term is physically justified, as it reflects probing of unoccupied d-states via p-d hybridization. For Cu, this feature arises from the specific shape of the unoccupied p-state density of states and the absence of unoccupied 3d-states. However, since d-state occupancy changes under compression (as demonstrated later), we denote all such absorption edge kinks as "pre-edge" features for clarity.

A key phenomenon at the Zr K-edge (FIG. 1(b)) is the systematic increase in the intensity of the pre-edge peak. In a classical framework, 1s→4d transitions are dipole-forbidden (Δl=2). We interpret the enhancement of this signal under pressure as a direct consequence of increasing p–d orbital hybridization. Compression forces the overlap of the spatially extended 5p orbitals with the more compact 4d orbitals, thereby introducing the d-character states with a p-character admixture and rendering the transition "partially allowed". In amorphous systems, where local inversion symmetry is broken, this effect is further amplified. This effect is seen in a piecewise-linear fit to the Zr pre-edge intensity, which identifies a breakpoint at 44(7) GPa (FIG. 1(d)). The total change in the Zr pre-edge intensity $\Delta\mu_{pre-edge} = 0.27$ (taken from normalized $\mu(E)$ – see normalization procedure in SM). Similarly, a piecewise-linear fit to the Cu pre-edge height identifies a breakpoint at 30(8) GPa, consistent with the DFT calculations shown below. In the case of Cu, the total change in pre-edge intensity $\Delta\mu_{pre-edge} = 0.07$. Additionally, the white-line heights do not change significantly. For Zr, the white-line intensity remains essentially invariant across the entire pressure range, whereas for Cu, a systematic increase is observed, reaching $\Delta\mu = 0.05$.

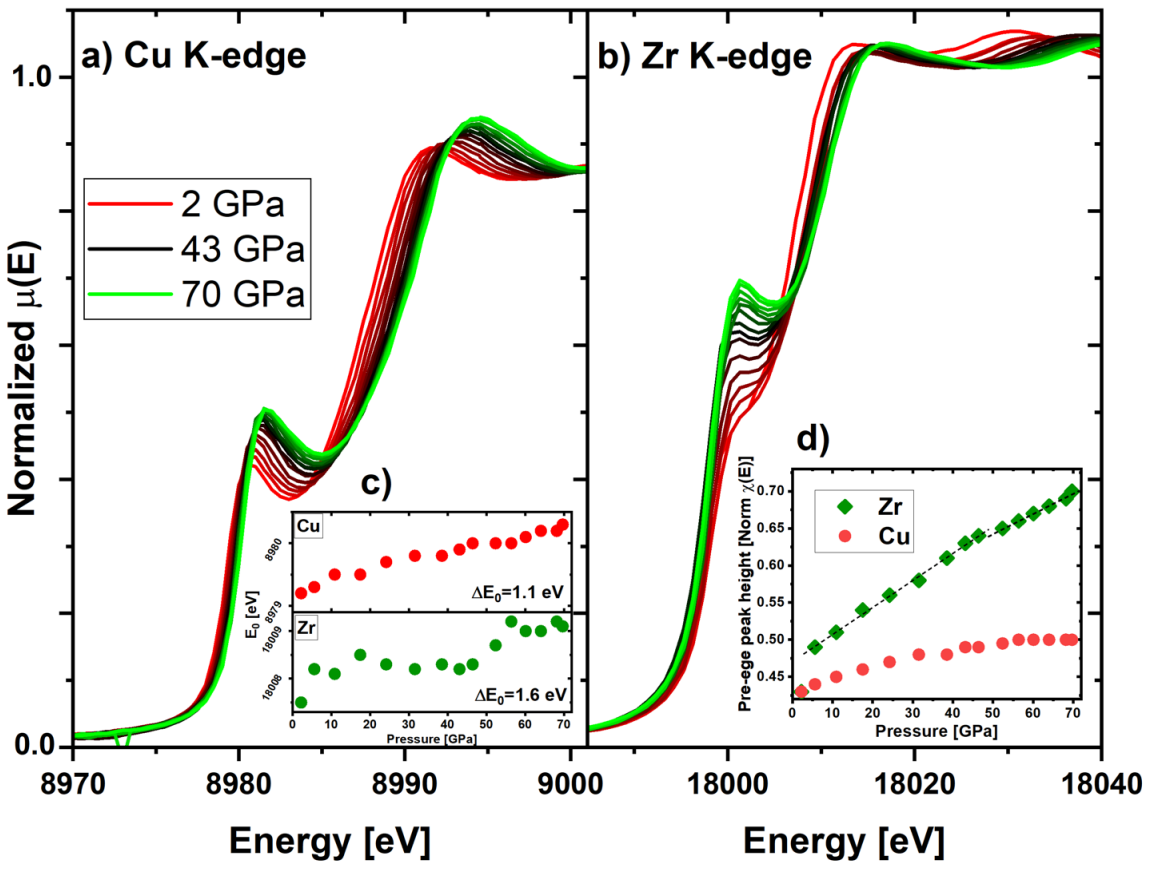

FIG. 1. Experimental XANES spectra for (a) Cu and (b) Zr absorption edges across a wide range of pressures. Insets (c) and (d) show the evolution of the absorption energy and normalized Zr and Cu pre-edge

*Contact author: przemyslaw.dziegielewski@pw.edu.pl

intensities, respectively. Zr pre-edge height exhibits two distinct linear regimes with a crossover at 44 GPa. Dashed lines are guides to the eye.

### B. MD data validation

Molecular dynamics simulations provide essential insights into the thermodynamic and structural properties of the investigated glass. While the material properties and compression-induced effects were discussed in our previous studies [11,32,34,35], they lacked sufficient experimental validation of the employed interatomic potential. Using our current experimental results, we compared EXAFS spectra derived from MD-generated structures (calculated with the FEFF code [36,37]) with the measured data (see the SM for details). FIG. 2 presents the EXAFS spectra for both absorption edges at three representative pressures: 10 GPa (low), 50 GPa (intermediate), and 70 GPa (high). The MD-derived EXAFS spectra reproduce both the oscillation frequency and phase, indicating that the dominant interatomic distances are correctly represented.

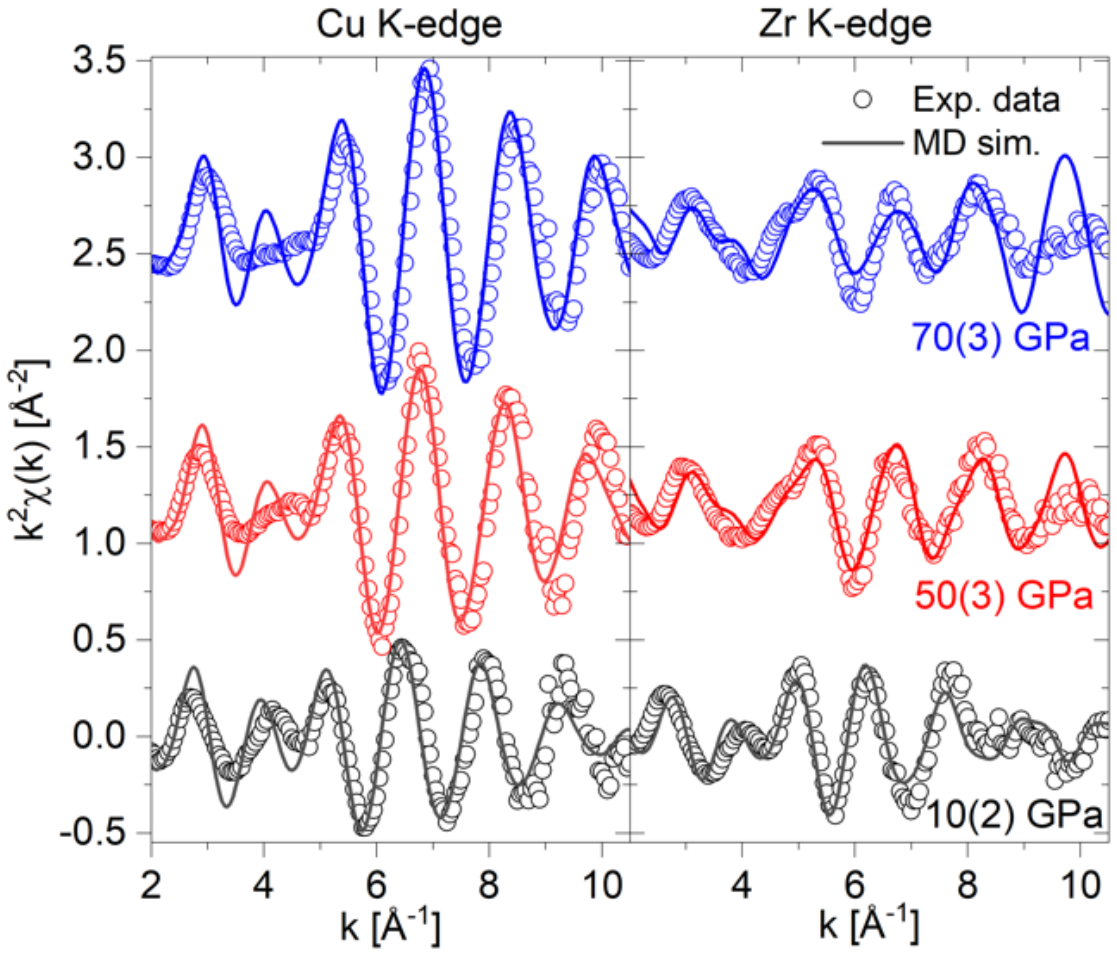

FIG. 2. Experimental EXAFS function (points) measured at three pressures: 10(2), 50(3), and 70(3) GPa, and calculated EXAFS functions (lines) based on atomic configurations from MD simulations.

Since EXAFS in amorphous alloys is primarily sensitive to the first coordination shell, these results confirm that the simulations effectively capture the local short-range order. The correspondence in oscillation amplitudes is also noteworthy, suggesting an accurate reproduction of the coordination numbers. Notably, the EXAFS oscillation amplitude at the Cu K-edge increases with pressure, whereas it remains nearly constant at the Zr K-edge. This trend aligns with previous studies indicating that compression drives the structural ordering of Zr–Cu alloys into Cu-centered icosahedral clusters, a process that occurs at the expense of the local order around the Zr atoms. Quantitative analysis yields a root-mean-square error (RMSE) below 0.1 $Å^{-2}$ across the entire pressure range, which represents an acceptable threshold for modeling amorphous systems. This comparison confirms that the simulation procedure and the embedded-atom-method (EAM) potential employed accurately represent the experimental system, supporting the use of MD-generated configurations as reliable input for subsequent DFT calculations.

Despite the fact that MD cooling rates are several orders of magnitude higher than experimental ones - a factor known to affect the resulting glass structure [38,39] - the high degree of correspondence with EXAFS supports the reliability of our approach at the level of local atomic order. The qualitative agreement between the experimental and computational data is satisfactory.

### C. Charge transfer – DFT analysis

Describing the electronic structure of MGs is challenging due to the absence of a periodic lattice, which makes computations costly. In MGs, metallic bonding dominates, and the valence electrons exhibit nearly free-electron-like behavior [40]. Electronic delocalization leads to a shared electron cloud in which immobile ions are embedded, with wave functions extending well beyond typical interatomic distances. Consequently, a single electron is shared among many atoms, leading to strong orbital hybridization. Upon compression and the associated reduction in interatomic separations, wave-function overlap increases, and electrons closer to the atomic core become increasingly important. Due to strong hybridization, orbital populations should be interpreted qualitatively rather than as absolute charges.

Charge-population analysis using the Mulliken scheme [41,42] effectively tracks electronic changes occurring in the present system. This analysis yields orbital-resolved populations that can be attributed (within the limitations of the method) to specific atomic states. Hereafter, Δq represents the net charge change; positive values signify an electron gain, while negative values signify a loss (FIG. 3(a)).

*Contact author: przemyslaw.dziegielewski@pw.edu.pl

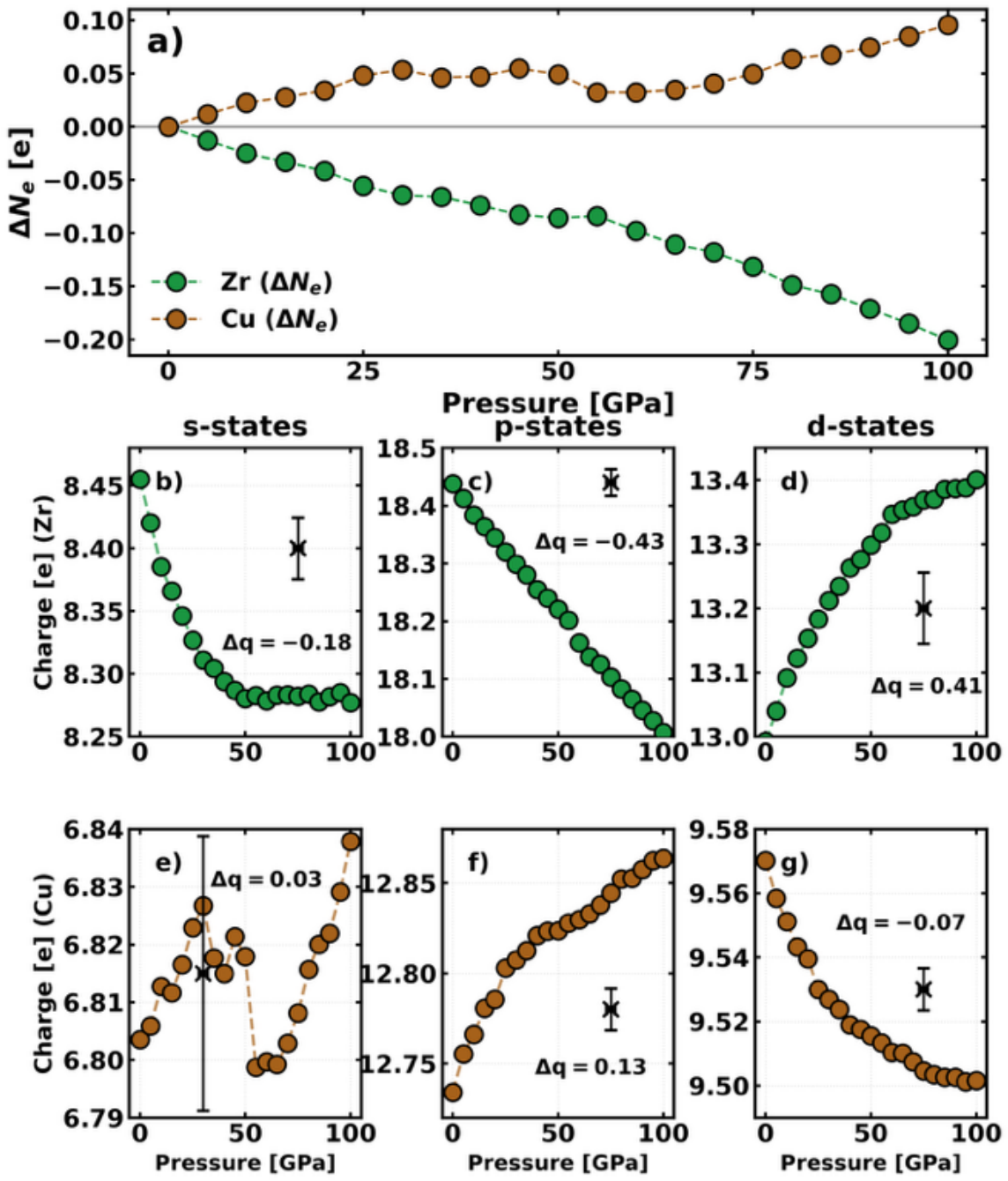

FIG. 3 (a) Net charge changes for Cu and Zr, and average Mulliken populations for (b–d) Zr and (e–g) Cu s, p, and d-states with the charge uncertainties. Simulation data are shown up to 100 GPa, extending beyond the experimental range of 70 GPa.

Across the 0-100 GPa range, charge is transferred from Zr (Δq=-0.13e) to Cu (Δq=0.25e). This transfer is twice as large for Cu because the system contains twice as many Zr atoms as Cu atoms. In FIG. 3(b), a pronounced decrease in the Zr s-electron population with pressure is observed. The change amounts to Δq=-0.18e (electron charge) with a breakpoint at 32(3) GPa and a plateau above 50 GPa (see bootstrap analysis in SM). This provides a direct quantification of the depopulation of the 5s orbital that we postulated earlier. The change in the Zr p-states (approximately Δq=-0.43e) is approximately linear, accompanied by a step-like drop above 50 GPa (FIG. 3(c)). A formal breakpoint fit places this feature near 76 GPa, but the bootstrap distribution for this channel is broad and multimodal rather than sharply peaked, so we do not treat this as a well-resolved threshold (see SM). The increase in the Zr d-electron population (approximately Δq=0.41e) with a breakpoint at 48(5) GPa correlates with the reduction of charge in the s and p-states: the 4d band acts as an acceptor for the charge that leaves the s and p orbitals (FIG. 3(d)). The sum of the Zr s and p charge transfer does not equal the increase in the Zr d charge due to the presence of Cu and non-negligible charge transfer to Cu p-states and Cu-Zr electronic state hybridization.

Only minor changes, an order-of-magnitude lower than for Zr, occur in Cu s-state population (Δq = 0.03e) (FIG. 3(e)). Electrons from Cu p-states (Δq=0.13e, breakpoint at 26(3) GPa) are partially exchanged with the Cu d-states (Δq=-0.07e, breakpoint 32(4) GPa) (FIG. 3(f-g)). The changes do not sum to zero - the uncompensated charge is the portion shared between Zr and Cu. This allows us to conclude that the charge redistribution occurs primarily within the atomic region of a given element, with only a small Zr–Cu charge transfer. In addition, two regimes can be distinguished in the pressure dependence of the Mulliken charges for Zr (see FIG. 3 (b-d)), with the boundary between these regimes falling in the range 30-60 GPa. However, the Mulliken populations quantify orbital-resolved charge (in units of electrons), whereas Rahm's prediction on an electronic reconfiguration at 54 GPa is expressed in terms of electronegativity, an energy-based quantity. These two descriptors are not directly comparable, and, in the following paragraphs, a more appropriate point of comparison is provided by the orbital-energy analysis.

The calculated density of states (DOS) shown in FIG. 4 provides complementary insight into the pressure-dependent electronic structure of the metallic glass. Three pronounced maxima in the total DOS indicate the presence of three main energy bands (FIG. 4 (a)). Although the total DOS is dominated by Zr and Cu d-states (FIG. 4(f,g)), the pressure evolution of the XANES spectra is governed primarily by subtler changes in the s and p-states (FIG. 4(b–e)). For Zr, the s- and p-projected DOS decreases markedly with increasing pressure, whereas the d band broadens and extends to higher energies above $E_F$. Because K-edge XANES probes unoccupied states with p character, these changes are directly relevant to the observed spectral evolution [32]. Experimentally, the Zr white-line intensity remains nearly constant over a broad pressure range, whereas the pre-edge intensity increases strongly. This behavior cannot be explained by the density of final states alone: the number of Zr d states near $E_F$ actually decreases under compression, particularly in the −2 to +2 eV range, despite the overall band broadening. Instead, the rising pre-edge intensity points to enhanced p-d hybridization. Although the direct 1s→4d/5d transition is dipole forbidden, hybridization transfers p character to a fraction of the d-like final states, thereby increasing the transition probability. In this way, the hybridization-driven increase in the matrix element outweighs the reduction in the number of available final states, consistent with the XANES observations.

*Contact author: przemyslaw.dziegielewski@pw.edu.pl

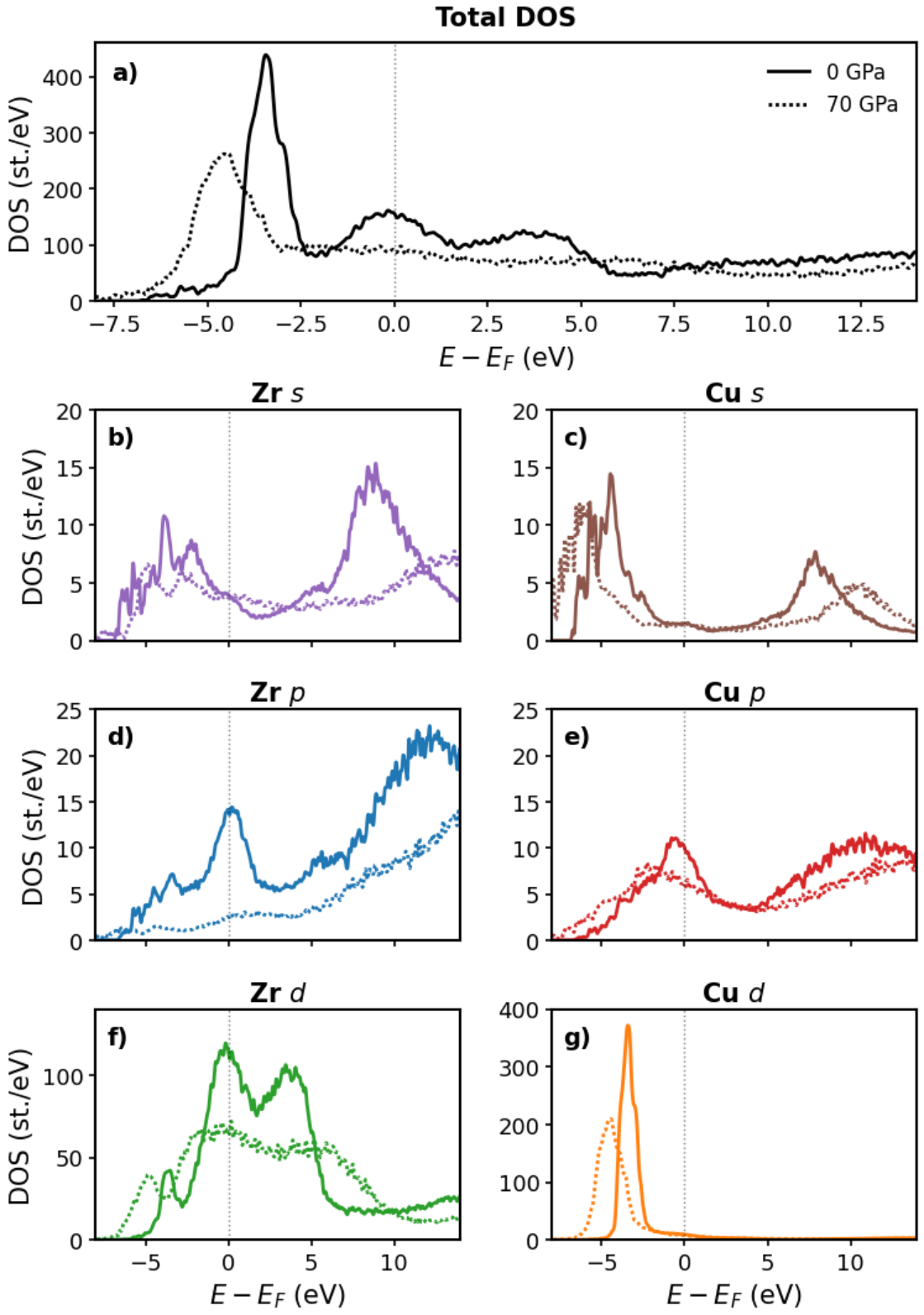

FIG. 4. Calculated total DOS (a) and partial DOSes for Zr (b, d, f) and Cu (c, e, g) s,p,d-states respectively at 0 GPa (solid lines) and 70 GPa (dashed lines).

In contrast, the Cu DOS changes only weakly with pressure. The number of unoccupied p-states is very similar at low and high pressure (FIG. 4(d)), so the modest enhancement of the Cu white line and pre-edge feature in XANES is induced by Cu-Zr hybridization, not by an increase in the number of empty p-states. Since the Cu 3d band is fully occupied, the weak evolution of the Cu pre-edge is also governed primarily by p-d hybridization. Importantly, this hybridization is not purely local to Cu, but arises from Zr-Cu orbital overlap; therefore, the pressure-induced electronic reconfiguration of Zr also modifies the Cu spectral signature. This interpretation is supported by the Mulliken population analysis, which shows that the Zr transformation leaves a measurable imprint on the Cu electronic environment.

To support raw Mulliken/DOS analysis with a quantity more directly comparable to Rahm’s energy-based prediction, the first moments (band centers) of the species and orbital-resolved density of states were computed $\epsilon_l = \frac{\int E n_l(E) dE}{\int n_l(E) dE}$, where $n_l$ is a density of states for energy $E$, and $l$ is the angular momentum quantum number. Band centers were calculated for the valence band up to $E_F$. This descriptor, widely used to characterize orbital energetics in the d-band model of Hammer and Norskov [43] and, more generally, grounded in the moments theorem of Cyrot-Lackmann for disordered tight-binding systems [44], provides an energy-resolved counterpart to the charge-based Mulliken analysis.

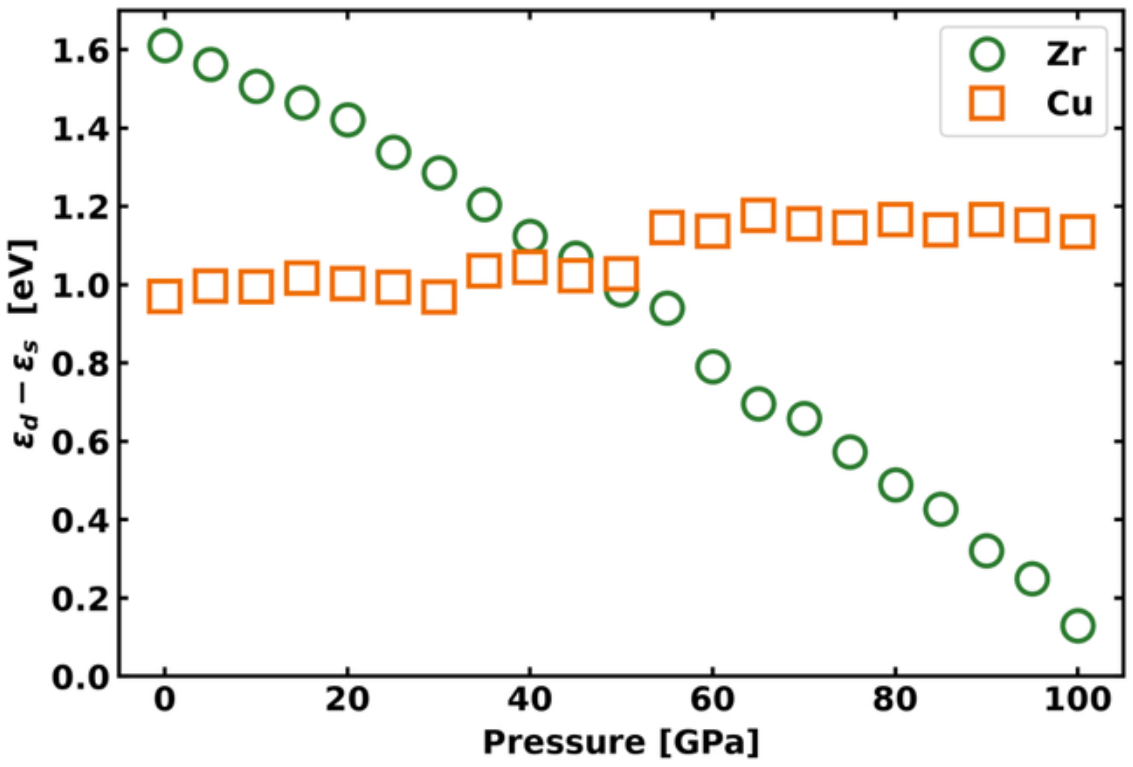

FIG. 5. Zr and Cu s-d band centers separation calculated up to $E_F$.

For Zr, the s-d orbital-energy separation decreases monotonically from 1.6 eV at ambient pressure to 0.13 eV at 100 GPa, reflecting the progressive energetic convergence of the s and d-states under compression (FIG. 5). This effect is qualitatively consistent with the direction of Rahm's predicted reconfiguration. In the case of Cu, the s-d orbital-energy separation is almost constant (~1.0 eV) up to about 55 GPa, and then increases slightly to ~1.15 eV at 100 GPa - the opposite trend to Zr, where the s-d gap narrows under compression. This contrast is consistent with the comparatively minor electronic redistribution observed for Cu throughout this work. The Cu 3d band is already fully occupied, compression does not drive an analogous s→d transfer, and the slight widening of the Cu s-d separation reflects a much weaker jump rather than a signature of orbital reconfiguration.

As the Zr p-states lose a number of occupied states (FIG. 4(d)), but the band center does not simply disappear - it genuinely shifts upward, above $E_F$, consistent with the general effect by which compression pushes orbital energy levels across the Fermi level [45]. Because so much of the p-band weight now sits above $E_F$, an occupied-only band center becomes unreliable at high pressure. This differs from the semicore-band broadening observed for pressure-activated alkali-metal p orbitals [45], where the band remains in place but widens; here, the p-band instead shifts entirely away from the occupied states.

*Contact author: przemyslaw.dziegielewski@pw.edu.pl

Three quantities - the Zr s-state Mulliken population, the Cu d-state Mulliken population, and the experimentally measured Cu pre-edge intensities - converge on a consistent lower pressure of approximately 30 GPa. A second crossover close to 50 GPa is identified independently by two complementary probes of the Zr d-orbital occupancy: the Zr d-state Mulliken population (48 GPa) and the Zr pre-edge XANES intensity (44 GPa) with 5 GPa uncertainty. We interpret this second crossover as reflecting the completion of d-band charge accumulation and its associated spectroscopic signature, whereas the lower-pressure 30 GPa crossover reflects the initial orbital-energy reorganization captured by the earliest charge-transfer response.

The higher-pressure breakpoints (44-48 GPa) identified independently by the Zr d-state population and from the Zr pre-edge intensity lie close to, though still below, the 54 GPa threshold predicted for the isolated Zr atom [2]. The lower-pressure breakpoints (around 30 GPa) show that electronic reconfiguration begins at a substantially lower pressure than this atomic-limit value. This two-stage picture is compatible with the 50 GPa threshold reported for the anomalous splitting of the first Zr-Zr coordination shell in prior MD studies [11–14]. That threshold reflects the macroscopic pressure applied to the sample. The local pressure and coordination environment of individual Zr atoms in a topologically disordered glass varies substantially from atom to atom. A subset of atoms may therefore undergo electronic reconfiguration and its structural consequences at a local pressure different from the global average. This could account for the gap between the two breakpoints identified here.

For Zr, the correlated DFT and XANES results confirm that the pressure-induced changes are consistent with an s,p→d redistribution in Zr, including 5s depopulation and increased d-band participation. Changes in the p-state DOS and p-d hybridization explain the approximately constant white-line intensity together with the increasing pre-edge intensity observed in XANES. For Cu, the data confirm that the 3d-states form a closed, fully occupied band, and that the observed evolution is driven by Zr-Cu hybridization, which accounts for the weaker pressure dependence of the pre-edge feature.

## IV. CONCLUSIONS

We have shown that compression of amorphous $Zr_{67}Cu_{33}$ metallic glass induces a pronounced redistribution of Zr valence electrons from s and p-states toward d-states. The primary experimental signature is the strong growth of the Zr K-edge pre-edge feature, with a breakpoint near 44 GPa. This behavior is consistent with the DFT results, which show decreasing Zr s- and p-state populations, increasing d-state population, and enhanced p-d hybridization under pressure. In contrast, Cu shows only modest electronic changes, indicating that the dominant transformation centers on zirconium. Notably, the redistribution involves not only the 5s depopulation predicted by Rahm's model but also significant p→d transfer, reflecting the role of orbital hybridization in the condensed multicomponent system.

The agreement between experiment, molecular dynamics, and electronic-structure calculations supports the interpretation that the anomalous high-pressure evolution of this metallic glass is not solely governed by structural packing but is closely tied to a pressure-driven reorganization of the Zr electronic states. The 44-48 GPa breakpoints identified here for the Zr d-orbital occupancy lie close to the 54 GPa threshold predicted for the isolated Zr atoms. The 30 GPa breakpoints mark the onset of electronic reconfiguration, indicating that the metallic glass environment favors it at lower pressure than the atomic limit. This finding still provides insights into pressure-driven transformations in amorphous metallic systems, and suggests that electronic reconfiguration may be a general driver of polyamorphic transitions in metallic glasses.

## ACKNOWLEDGMENTS

We acknowledge Diamond Light Source for time on I18 under proposal SP32715-1 and thank Dr Konstantin Ignatius for assistance with the beamline. This research was funded in whole or in part by the National Science Centre, Poland, Miniatura grant number 2023/07/X/ST3/01224. This research was carried out with the support of the Interdisciplinary Centre for Mathematical and Computational Modelling at the University of Warsaw (ICM UW) computational grants g98-2072 and g103-2505. This work was supported by the National Science Centre, Poland, grant agreement No 2021/43/B/ST5/02480. OTL acknowledges support from the Royal Society in the form of a University Research Fellowship (UF150057).

*Contact author: przemyslaw.dziegielewski@pw.edu.pl

*Contact author: przemyslaw.dziegielewski@pw.edu.pl

*Contact author: przemyslaw.dziegielewski@pw.edu.pl

Supplementary information for “Pressure-induced s,p-d electron redistribution accompanies structural transformation in amorphous Zr-Cu alloy under compression”

Przemyslaw Dziegielewski*[1], Jerzy Antonowicz[1], Konstantinos Georgarakis[2], Oliver Lord[3], Monica Amboage[4], Dominik Daisenberger[4], Lewis Clough[5], Tetsuo Irifune[6], Toru Shinmei[6]

[1]Warsaw University of Technology Faculty of Physics Warsaw PL, [2]Cranfield University Faculty of Engineering and Applied Science Cranfield GB, [3]The University of Bristol School of Earth Sciences Bristol GB, [4]Diamond Light Source Didcot GB, [5]University of Edinburgh Edinburgh GB, [6]Ehime University Matsuyama JP

*przemyslaw.dziegielewski@pw.edu.pl

1. Rahm’s Theory predictions

Observed variation of the electronic structure can be explained by Rahm’s theoretical studies of isolated atoms under high pressure [1,2]. Rahm shows that pressure leads to changes in atomic radius, a decrease in electronegativity, and, for many elements, abrupt changes in electronic configuration. For Zr atoms with an electronic configuration of $5s^2 4d^2$, two pressure-induced reconfigurations occur: the first takes place at 2 GPa and involves the transfer of one electron from the 5s shell to 4d states (Fig. 1). The second transition occurs at 54 GPa and is associated with transfer of an electron from 5s to 4d states and the disappearance of the 5s orbital. At similar pressures, MD simulations show the formation of shortened Zr pairs. Additional consequences of these transitions include changes in electronegativity $\Delta\chi \approx 3.5$ eV/e and in the van der Waals radius $\Delta r \approx 0.08$ Å at the transformation pressure.

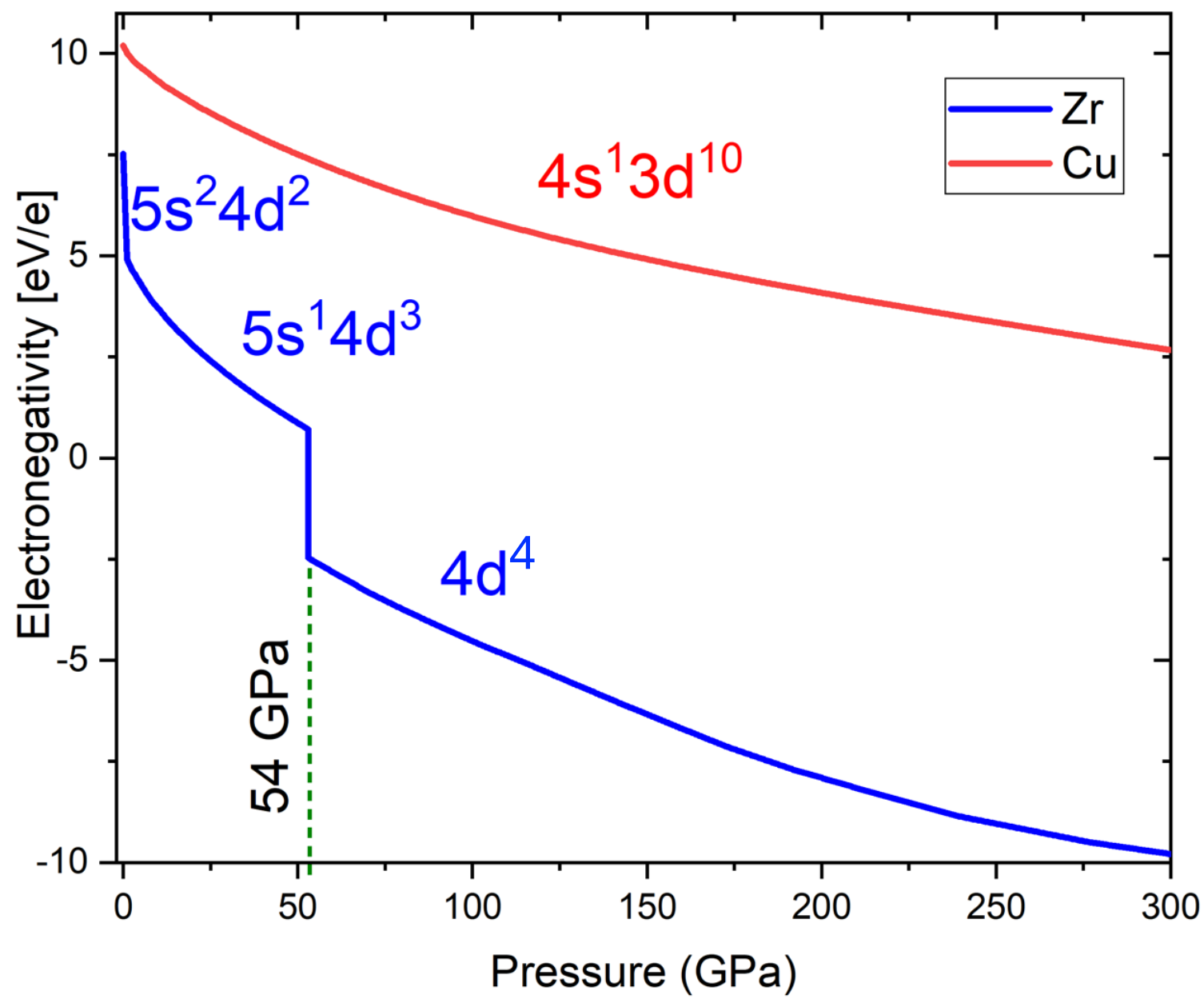

*Figure 1 Pressure-induced changes in the electronegativity and electronic configuration of an isolated atom according to Rahm’s theory. A characteristic pressure of 54 GPa drives an electronic reconfiguration of Zr, leading to the depopulation of the 5s orbital. Based on data from the "Atoms Under Pressure" project [3].*

2. Molecular dynamics

Molecular dynamics (MD) simulations were performed using LAMMPS [4] for a $Zr_{67}Cu_{33}$ system containing 43904 atoms, placed in a cubic simulation box with initial dimensions of 9.4×9.4×9.4 nm and periodic boundary conditions at 0 GPa and approx. 8x8x8 nm at 100 GPa. The simulations were carried out in the NPT ensemble using a Nosé–Hoover thermostat-barostat to control temperature and pressure, with a time step of 1 fs. The embedded-atom method (EAM) potential developed by Sheng was used [5]. The initial crystalline configuration taken from the Materials Project database was first pre-relaxed at 300 K and then heated over 1 ns from 300 K to 2000 K, approximately 700 K above the crystal melting temperature [6]. The system was equilibrated at 2000 K for an additional 1 ns and then quenched to 300 K at a cooling rate of $10^{12}$ K/s. The resulting glass was subsequently compressed up to 100 GPa at a rate of 0.1 GPa/ps. Using the same protocol, a classical MD simulation was also performed on a 128-atom system, which served as the initial (ambient) configuration for the DFT calculations. This smaller system size was sufficient to prevent self-interaction. The box size was approximately 13.5 Å, whereas the interaction cutoff used in the EAM potentials was 6.5 Å, smaller than half the simulation box dimension. Representative structures are shown in Fig. 2. Additionally, Fig. 3 shows the pair distribution functions (PDFs), g(r), calculated for both the large and small systems to illustrate the interatomic distances obtained using the OVITO Python library [7]. The additional peak originates from short-distance interatomic pairs, as discussed in our previous work [8,9].

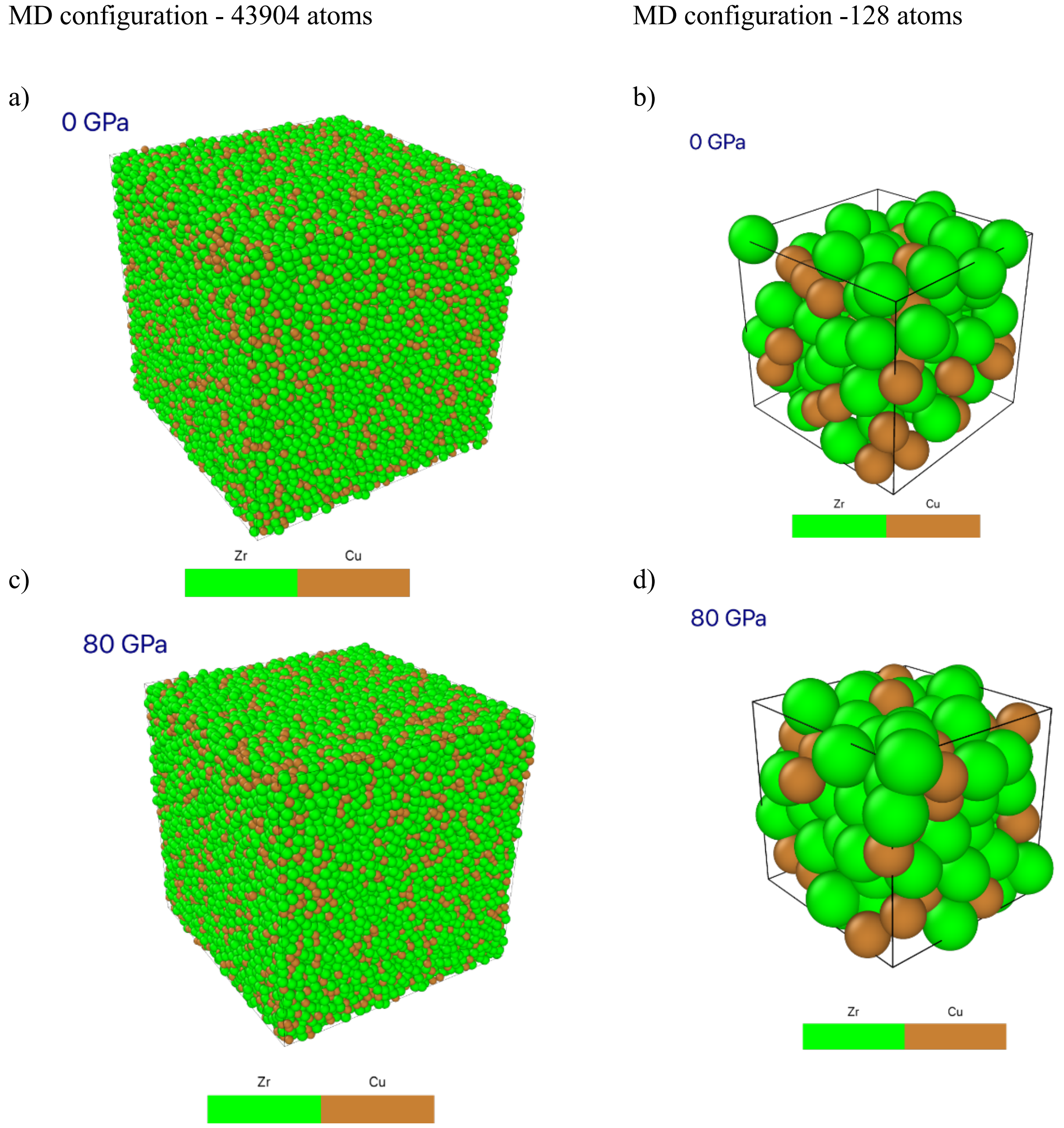

*Figure 2 Snapshots of molecular dynamics configurations for the large 43,904-atom system at (a) 0 GPa and (c) 80 GPa, compared with the corresponding configurations for the small 128-atom system at (b) 0 GPa and (d) 80 GPa.*

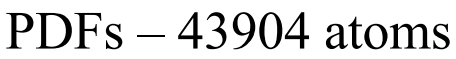

PDFs – 128 atoms

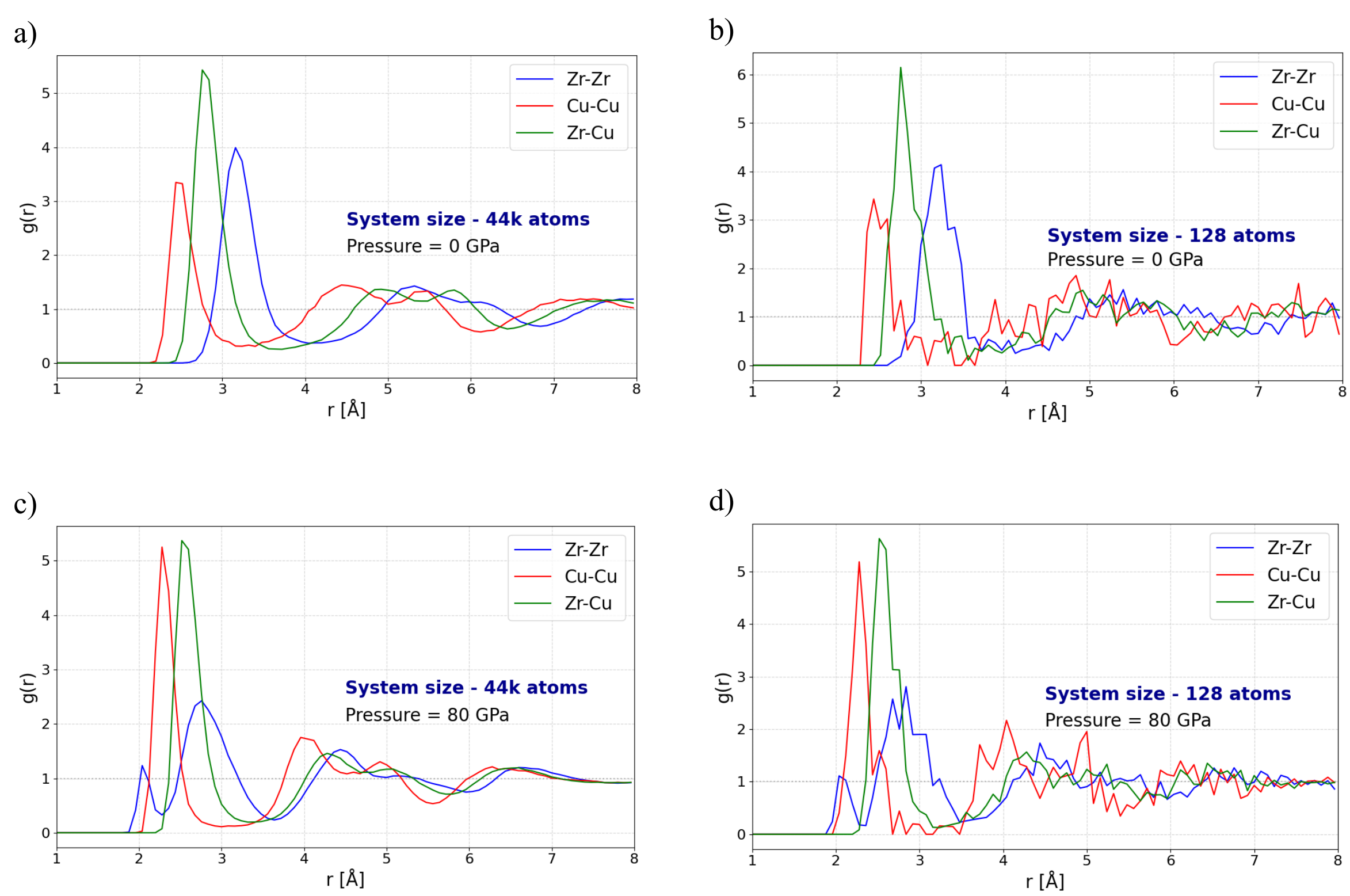

*Figure 3 Partial PDFs calculated from MD simulations for 0 and 80 GPa for a 43904-atom system (a) and (c), and the same for 128 atoms (b) and (d).*

## 3. DFT calculations

DFT calculations were performed using the all-electron, full-potential electronic structure code FHI-aims with an atom-centered basis of numerical atomic orbitals. Configurations obtained from MD simulations were used to calculate the system's electronic properties. The numerical integration and basis set construction followed the "intermediate" default settings provided by the package. We employed numeric atom-centered orbitals (NAO) with "tier 1" basis sets for both Zirconium (Zr) and Copper (Cu). For a) Zr: The minimal basis (5s,4p,4d) was augmented with a "first tier" of additional functions (hydro 4f, ionic 4d, ionic 5p, and ionic 5s). A 5g hydrogenic function was included as an auxiliary basis for multipole expansions. b) Cu: The minimal basis (4s,3p,3d) was augmented with the "first tier" set (ionic 4p, hydro 4f, hydro 3s, and hydro 3d), including a 5g function for auxiliary purposes. A dense integration grid was used, characterized by 434 points. The global species definitions included a potential cutoff of 4.0 Å, a trigger at 2.0 Å, and a decay of 1.0 Å, ensuring smooth convergence of the total energy. All calculations were performed with a 2×2×2 Monkhorst–Pack k-point grid and the PBEsol exchange-correlation functional [10]. Electronic-structure quantities (DOSes – Fig. 4, Mulliken charges) were calculated directly during DFT calculations. Mulliken charges [11] were averaged over the total number of Zr and Cu atoms. Charge-population analyses, such as the Mulliken scheme, are particularly informative for tracking the electronic changes occurring in the present system. Within this approach, each particle's contribution to the one-electron density matrix in Hilbert space is evaluated. The charge density at position $r$ is expressed in terms of the density matrix $P$ $\boldsymbol{\rho}(\boldsymbol{r}) = \sum_{\mu} \sum_{\nu} P_{\mu\nu} \phi_{\mu}(\boldsymbol{r}) \phi_{\mu}^{*}(\boldsymbol{r})$, where $\phi_{\mu}(\boldsymbol{r})$ are the basis functions used to represent the molecular orbitals. The Mulliken population, i.e., the net number of electrons associated with atom (k), is then obtained as $N_k = \sum_{\mu \in k} (PS)_{\mu\mu^*}$. The summation is restricted to basis functions centered on atom $k$, and $S$ denotes the overlap matrix. This procedure yields orbital-resolved populations that can be attributed (within the limitations of the method) to specific atomic states.

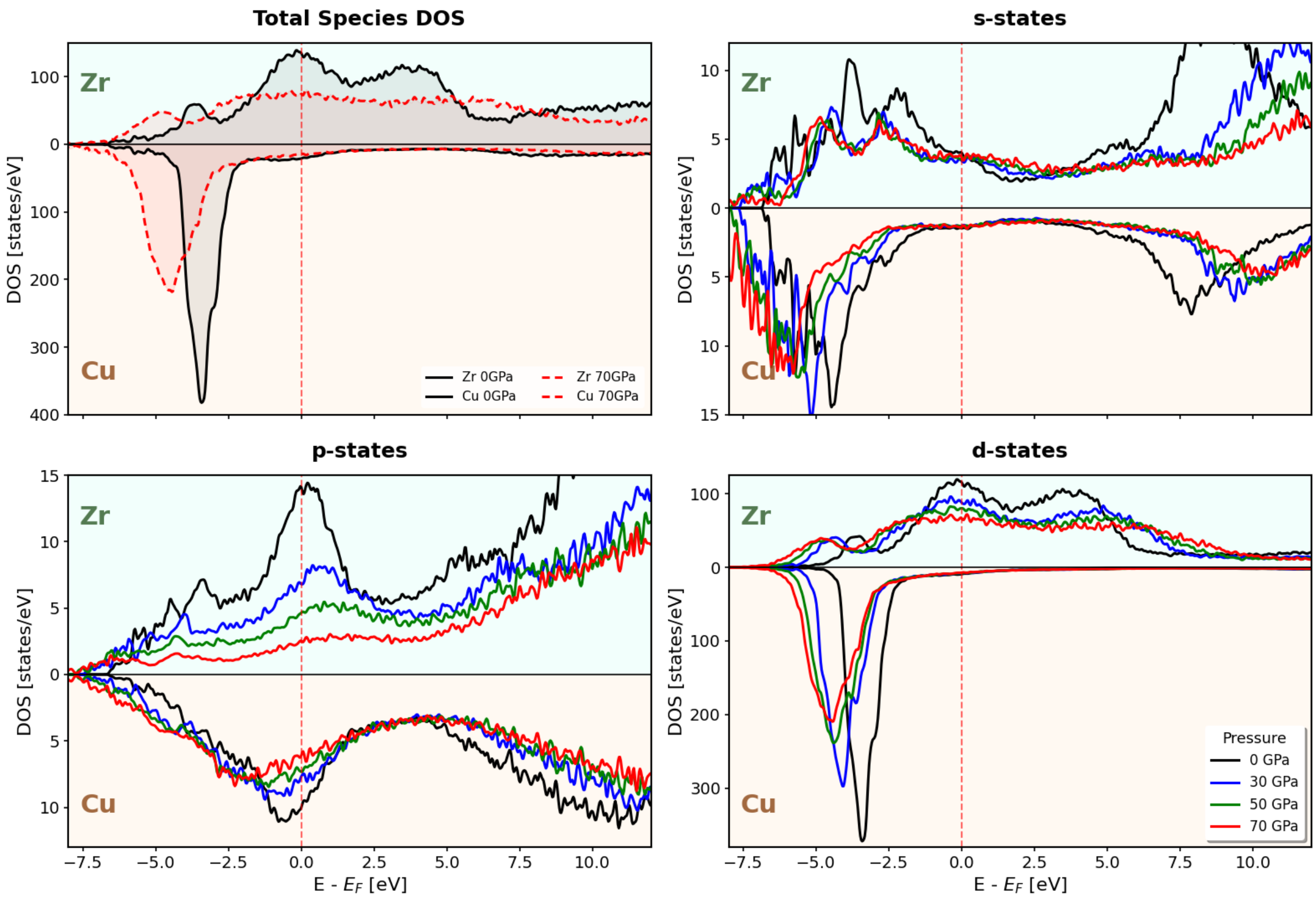

*Figure 4 Density of states and partial DOSes for Zr and Cu atoms in 0, 30, 50 and 70 GPa. To enable a more straightforward comparison, the DOS functions for both elements are presented in shared plots.*

4. Bootstrapping and breakpoint study

To test for pressure-induced changes in the slope of a given observable (band centers, Mulliken populations, and XANES pre-edge intensities), we fitted a continuous piecewise-linear model consisting of two linear segments that meet at a breakpoint, using ordinary least squares over a dense grid of candidate breakpoints. Confidence intervals for the breakpoint were obtained by residual bootstrap: the residuals of the best-fit model were resampled with replacement and added back to the fitted curve to generate 2000 synthetic data sets, each of which was refitted to obtain a bootstrap distribution of breakpoints; the reported 95% confidence intervals correspond to the 2.5th and 97.5th percentiles of this distribution [12,13]. This approach was chosen because it does not assume a particular sampling distribution for breakpoints and remains reliable for the small, irregularly spaced pressure grids characteristic of our DFT and experimental data sets. For regimes where the sum-of-squared-residuals profile decreases monotonically toward the edge of the fitting range rather than showing an interior minimum, no breakpoint is reported, as the data are better described by a single, smooth trend. For the Zr and Cu pre-edge XANES intensities, the lowest-pressure point (2.2 GPa) was excluded prior to fitting: at this point, the sample had only just been compressed within the diamond anvil cell, the applied stress was too small to produce a resolvable electronic response, and the associated pressure determination carries disproportionately large uncertainty. Retaining this point suppressed a genuine breakpoint in the Zr pre-edge; the resulting breakpoint (43.9 GPa) was independently confirmed using a robust (Huber-loss) fit to the complete, non-excluded data set, which converged to the same value, confirming that the result is not an artifact of the exclusion. Fitting curves, residual sum of squares profiles, and bootstrap histograms for all datasets discussed in the text, including this robustness check, are provided in Figures 5 and 6. Based on the analysis, Mulliken Cu-s charges were excluded due to high noise.

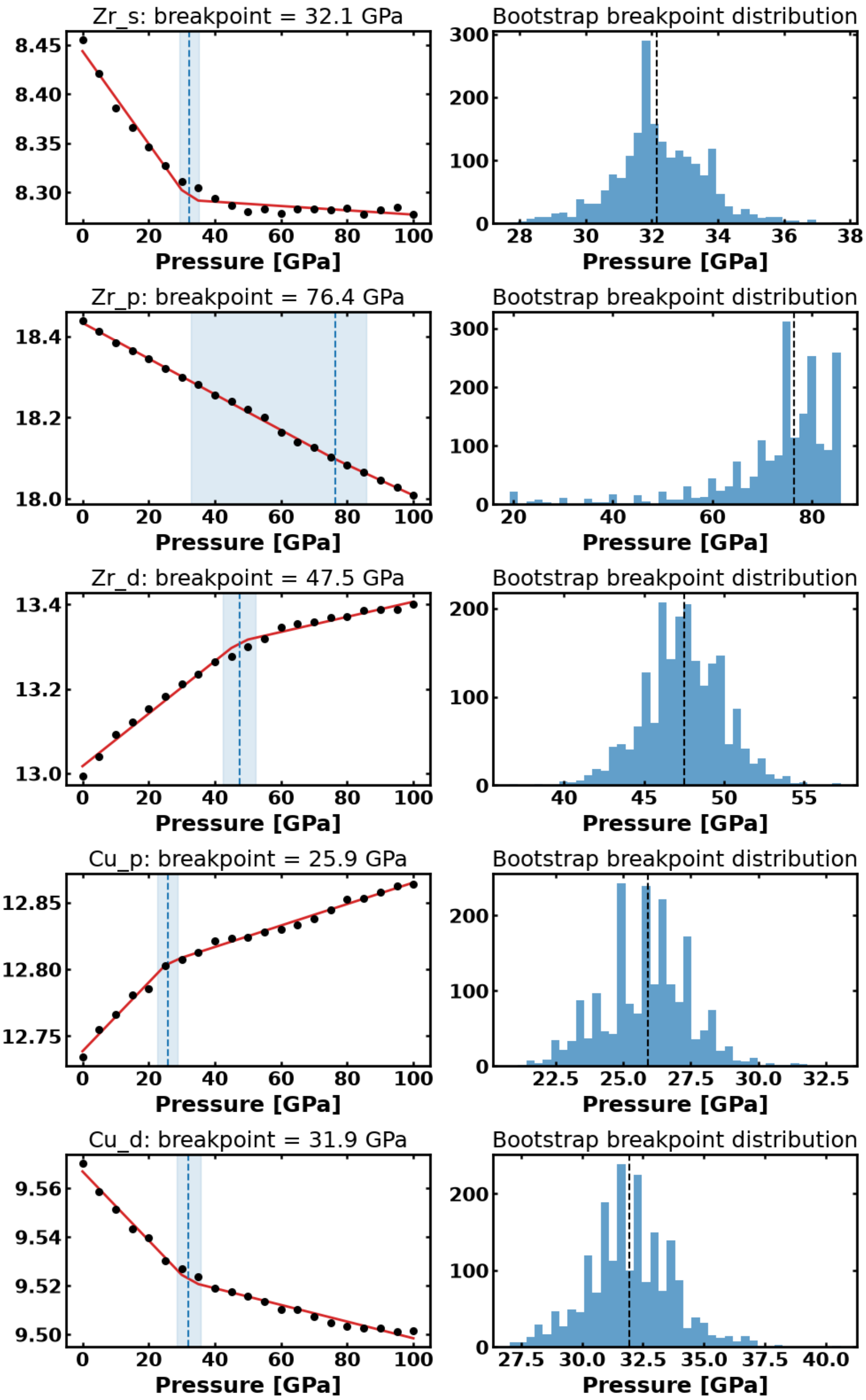

*Figure 5 Breakpoints with confidence intervals (left) and bootstrap breakpoint distribution (right) for Mulliken charges.*

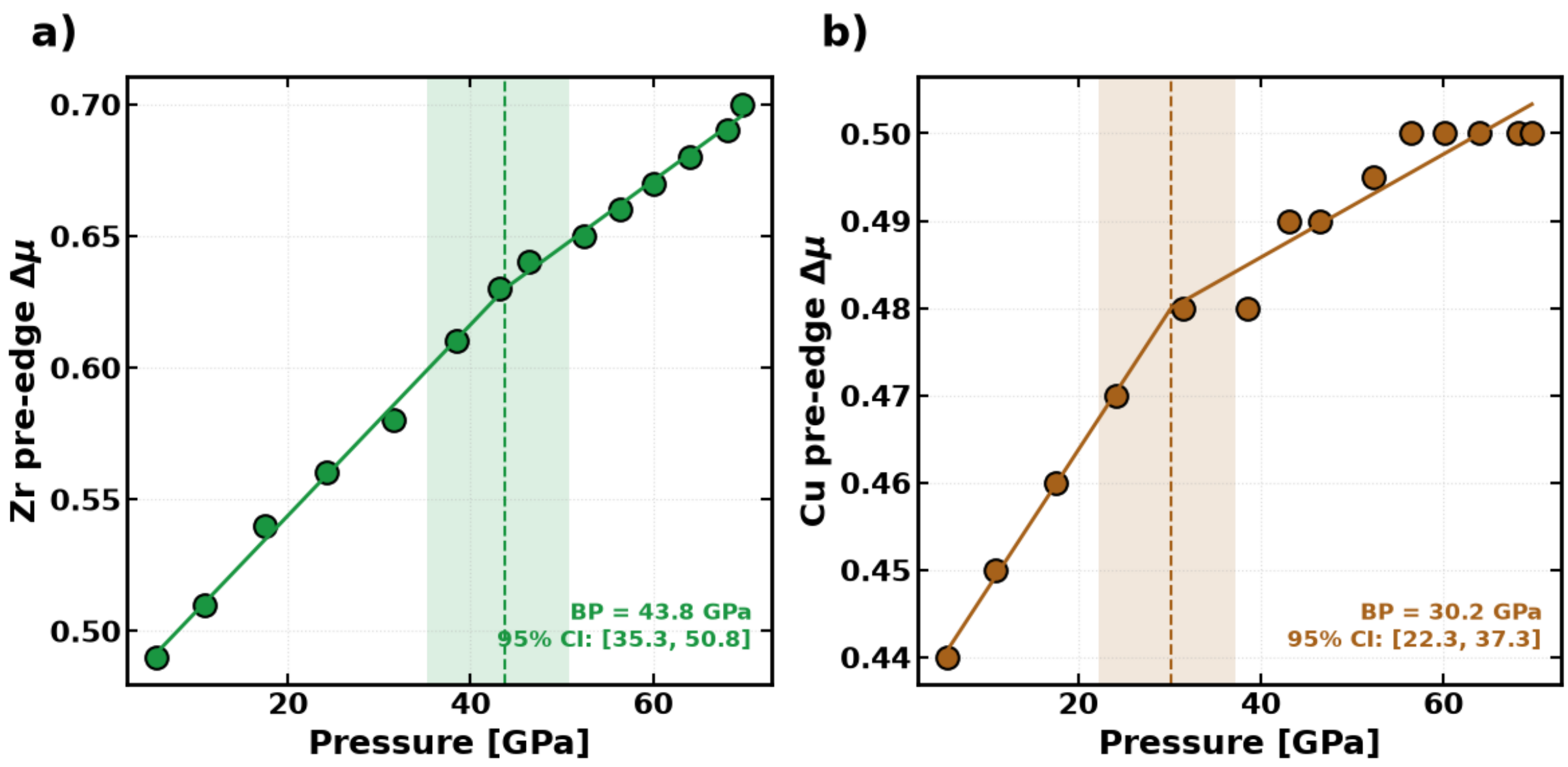

*Figure 6 XANES pre-edge breakpoints for a) Zr K-edge and b) Cu K-edge with confidence intervals.*

5. XAFS measurements

XAFS measurements were conducted at beamline I18 of Diamond Light Source. The $Zr_{67}Cu_{33}$ metallic glass foil was loaded in a symmetric diamond anvil cell (DAC) equipped with bevelled nano-polycrystalline diamond anvils (NPD) with culet diameters of 0.2 mm to prevent in EXAFS spectra due to the intense Bragg peaks produced from single crystal anvils. A sample chamber was drilled into a Re gasket pre-indented to a thickness of ~25 µm. A $Zr_{67}Cu_{33}$ metallic glass foil (~15 µm thick), a Au foil (~5 µm thick), and a single ruby sphere (~5 µm diameter) were placed into the chamber. The remaining space was filled with supercritical fluid Ne which acted as the pressure-transmitting medium, ensuring quasi-hydrostatic conditions. Pressure was increased in 15 steps up to 69 GPa. Reported pressures were measured by comparing the unit cell volume of the Au foil determined by XRD to its known equation of state [14], with an uncertainty of 1–3 GPa. The shift of the fluorescence peak of the ruby sphere, calibrated as a function of pressure, was used to determine pressure during initial compression. Both Cu (8979 eV) and Zr (17998 eV) K-edge spectra were measured at each P point. XRD was collected at 17800 eV without changing the experimental geometry, using an Excalibur detector. Before and after each XAFS measurement, an XRD map (Fig. 7) of the DAC was collected, enabling the X-ray beam to be positioned at the same location on the sample. In addition to the sample, reference spectra for metallic Zr and Cu foils at ambient conditions were measured for X-ray energy calibration.

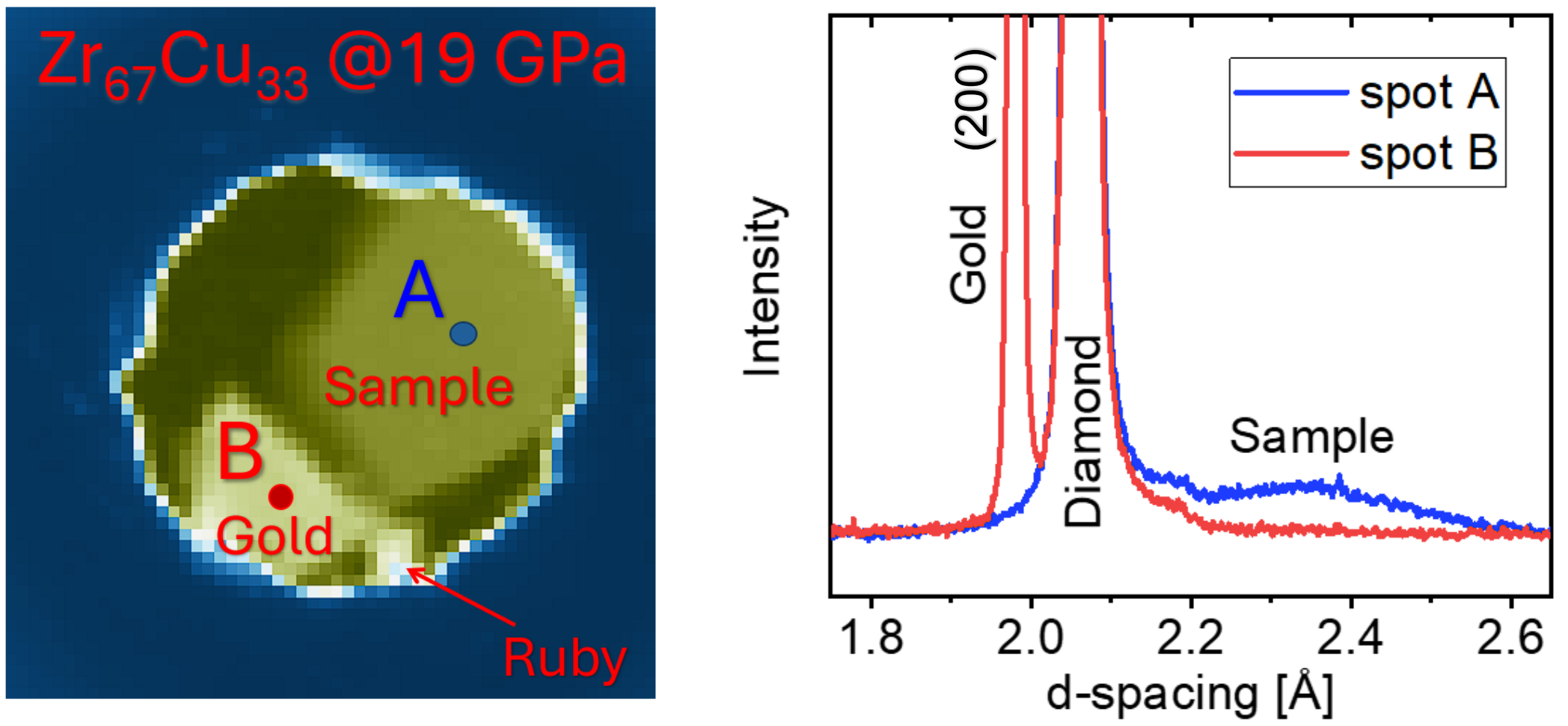

*Figure 7 Representative XRD map and diffraction pattern collected inside the DAC at a pressure of 19 GPa. The highlighted gold XRD peak indicates the (200) reflection.*

6. XAFS data normalization

Normalization of the XAFS data followed the standard protocol of fitting a linear pre-edge background (-100 to -30 eV relative to the absorption edge) and a post-edge polynomial (150 to 450 eV above the edge) (see Fig. 8). This normalization enabled a direct comparison of the pre-peak and white-line features. All data processing was conducted using the Athena software within the Demeter package [15]. Following normalization, the position and intensity of the pre-edge feature maximum were extracted. Absorption energy was determined by the position of the maximum of 1st derivative $\mu(E)$.

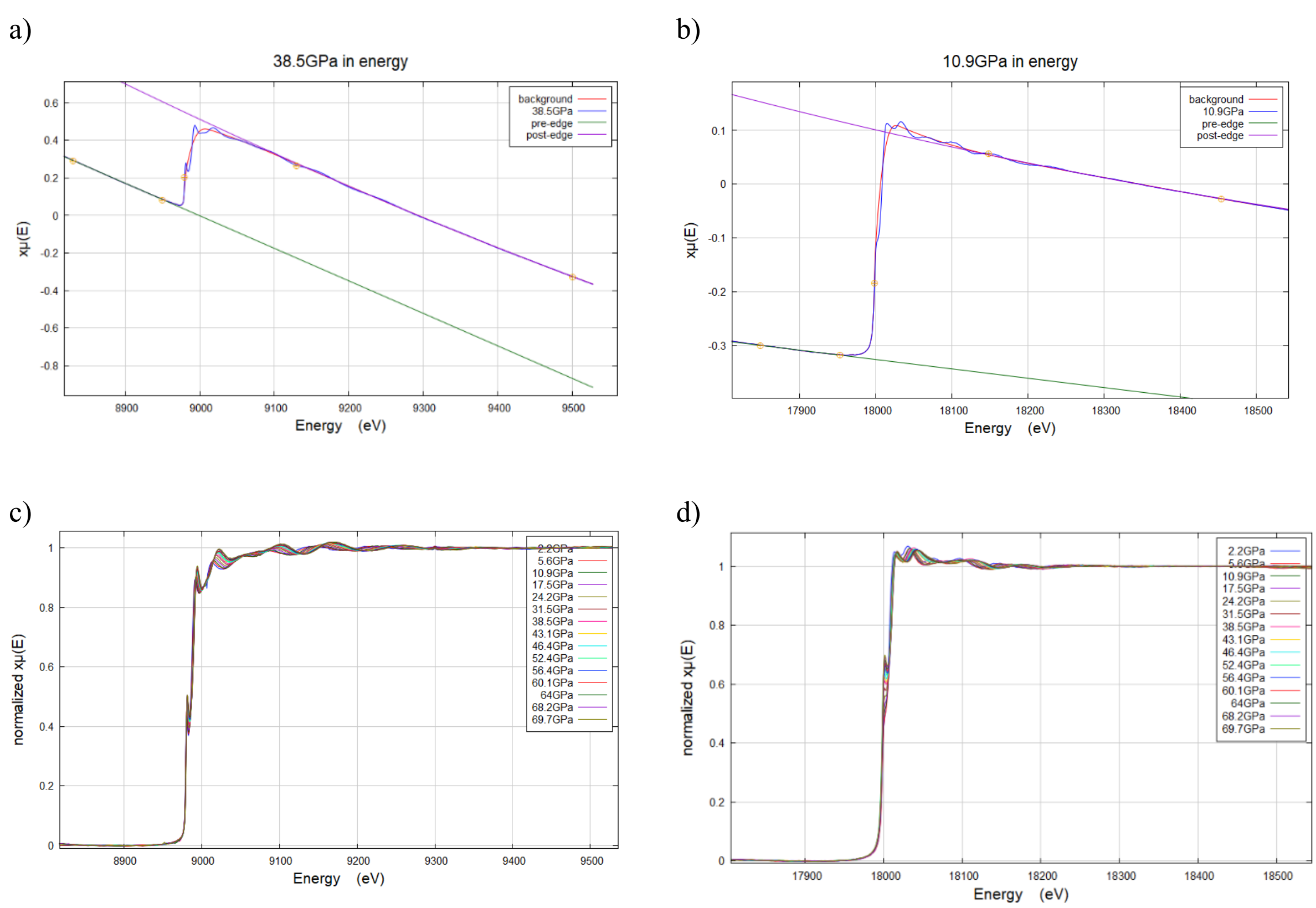

*Figure 8 Representative raw XAFS data for the (a) Cu and (b) Zr edges, including the fitted pre-edge and post-edge lines used for data normalization. Panels (c) and (d) present all measured XAFS spectra for the Cu and Zr edges, respectively.*

7. XANES and EXAFS calculations

To validate the simulations against experimental data, independent calculations were conducted for the XANES and EXAFS regions. XANES spectra were simulated using the FDMNES code [16] by averaging over all absorbing atoms (84 Zr and 44 Cu) within the simulation box and applying appropriate spectral broadening. Fig. 9 compares the MD-derived and experimental XANES spectra for the Zr and Cu edges under low and high pressures. These calculations spanned an energy range from −20 to +30 eV relative to the absorption edge, employing a 6 Å cluster radius and accounting for quadrupole contributions. EXAFS modeling was performed using FEFF10 [17]. For the MD trajectories, the EXAFS signal was averaged over 1000 randomly sampled absorbing atoms, considering scattering environments up to the third coordination shell. In contrast, the DFT-based EXAFS was averaged over all available absorbing atoms (84 Zr and 44 Cu). Fig. 10 presents a direct comparison of the calculated EXAFS for the larger and smaller systems.

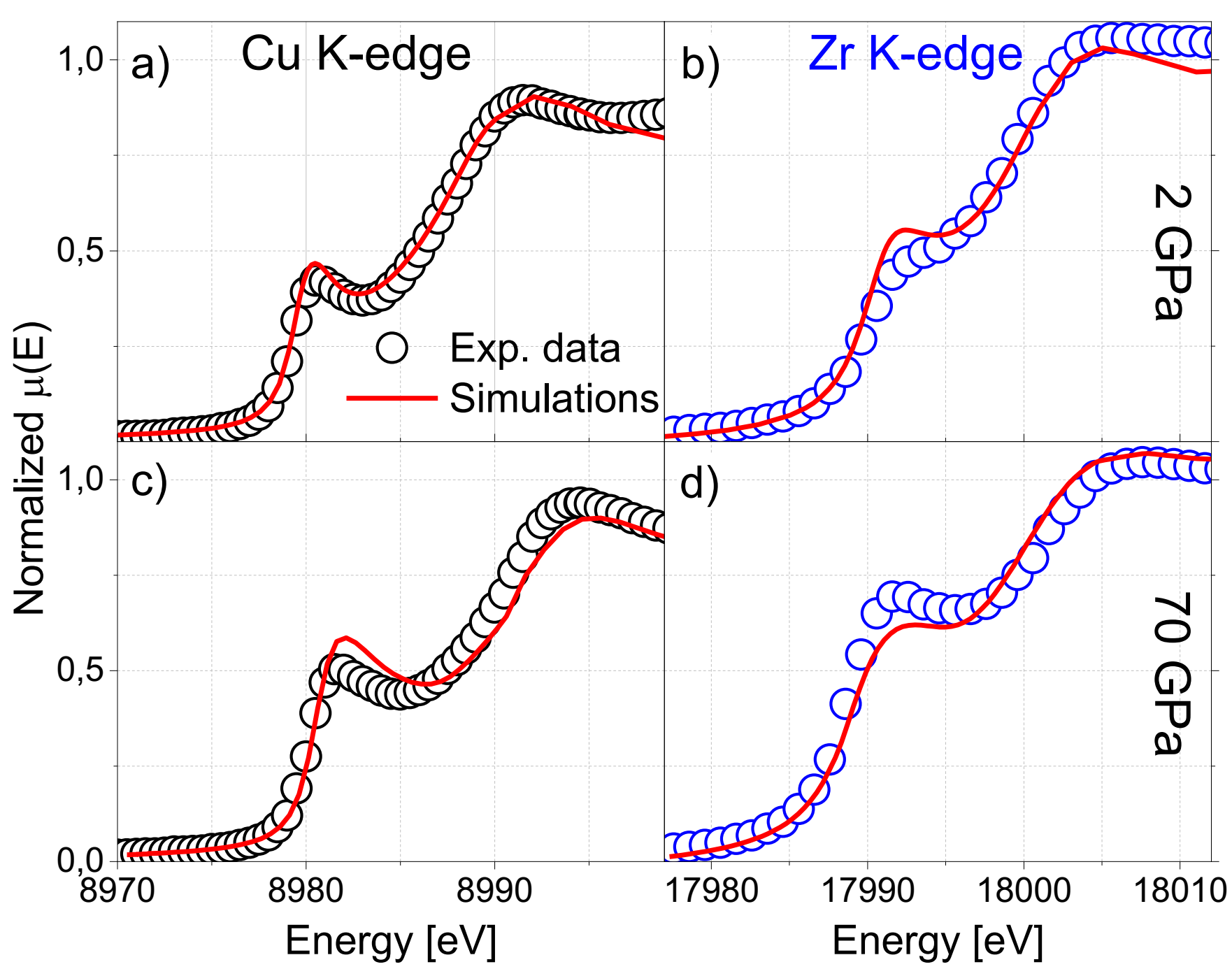

*Figure 9 Comparison of experimental data (points) and simulated spectra obtained using the FDMNES code for the Cu and Zr edges at low and high pressures.*

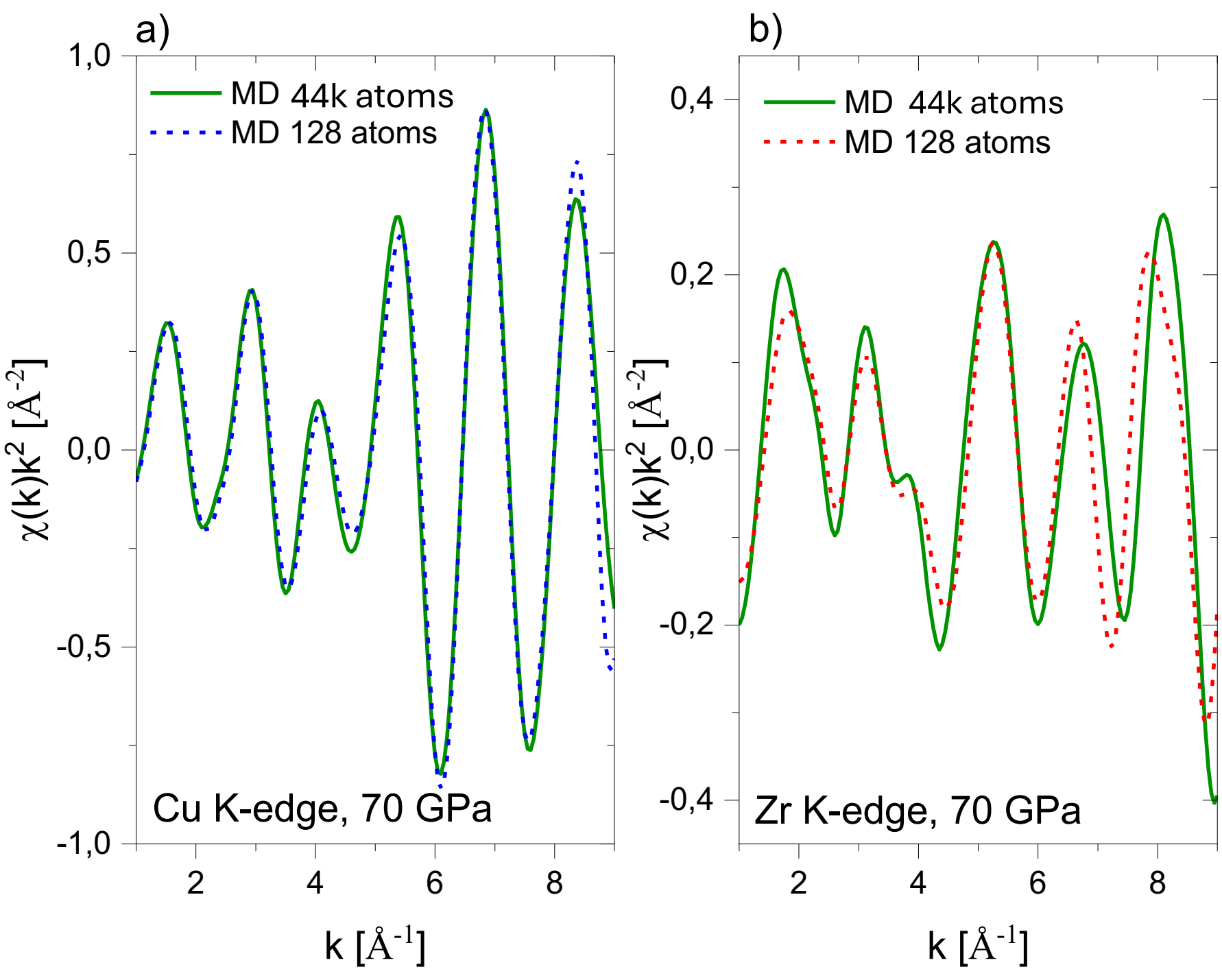

*Figure 10 Effect of the simulation box size and the number of atoms included in the EXAFS spectra calculations for a) Cu and b) Zr K-edge at 70 GPa.*